# Ion generation by photoionization and photofield tunneling of electrons during atom probe tomography of thermally grown chromia with deep UV laser light

Severin Jakob[1*], David Mayweg[1], Mattias Thuvander[1]

[1] Department of Physics, Chalmers University of Technology, SE-412 96 Göteborg, Sweden

*severin.jakob@chalmers.se

**Abstract**

The evaporation mechanisms during laser-assisted field evaporation of thermally grown chromia is investigated with the newest generation of commercial atom probe tomography instrument, equipped with a deep-UV laser (257.5 nm, 4.8 eV photon energy). By holding the voltage constant, the electrostatic field is kept constant, and the evolution of detection rate is recorded. The detection rate is measured as a function of laser pulse energy for chromia and the metallic alloy Ni-20Cr, where the expected thermal evaporation is observed. Furthermore, the detection rate as a function of electrostatic field is measured for chromia and follows the well-known Fowler-Nordheim type equation. The observation suggests that photoionization and photofield electron tunneling are the rate-limiting steps during evaporation of chromia with deep-UV laser light. The work function and bandgap of the material are discussed, and the evaporation behavior is put into context to existing observations.

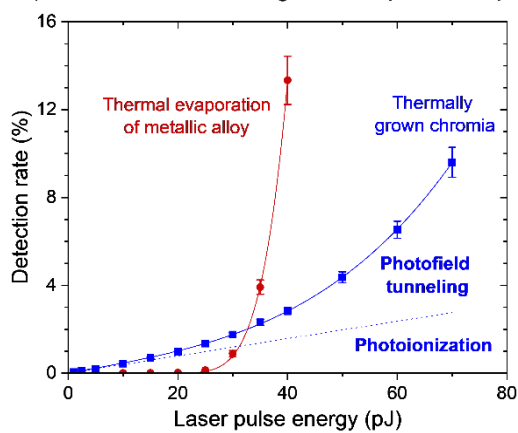

Atom probe tomography (APT) is a highly localized measurement technique, which can attain chemical information with almost atomic spatial resolution. The method is sensitive to all elements and is based on the field evaporation of single ions in a high electrostatic field. A needle-shaped specimen with an apex radius below 100 nm is exposed to high voltages, thereby reaching electrostatic fields in the range of tens of V/nm. For conducting materials an additional voltage pulse ionizes the atoms and accelerates them to a position-sensitive detector, allowing to identify the x- and y-position and furthermore the ion species by time-of-flight mass spectrometry. The temperature of the specimen is usually between 20 and 80 K. A reconstruction algorithm calculates the z-position of the ions from the measurement sequence [1]. Historically, the technique was limited to conductive materials, i.e. metals and alloys [2–4]. However, lasers with short pulses that enable a high energy density in the focal spot opened the technique to non-conductive materials [5–10]. Hereby, the voltage pulses are replaced by laser pulses, while the field is still created by the standing voltage. The widely accepted mechanism for field evaporation with laser light, for at least ultrashort (100 fs) pulses, is thermal activation [4,10]. In the case of metals, the electrons absorb the energy of the photons before transferring the energy to the lattice atoms [11,12]. The voltage to maintain a constant detection rate (DR) as a function of laser pulse energy (LPE) follows a negative linear trend for metals, whereas semiconductors with a small bandgap show a saturation effect with laser intensity [13,14]. The bandgap is expected to decrease in the electrostatic field and can even collapse, leading to a metallic behavior during field evaporation [15,16]. Conversely, in high bandgap semiconductors, vacancy defects at the sample surface are sufficient to allow strong laser absorption, followed by lateral and axial heating [17]. After the pulse, the tip is cooled through the shank and evaporation stops. Depending on the heat dissipation, peaks in the mass-spectrum show more or less thermal tails due to delayed cooling and delayed evaporation [18].

Beside thermal evaporation, photoionization of semiconductors has been observed. Tsong proposed three pathways for this process: i) direct ionization of an electron to the conduction band of the material, ii) excitation of the electron in the adatom and tunneling into the material, and iii) generation of a hole in the valence band by excitation from the photon in the material and tunneling of the electron from the adatom. His observations indicated pathway i) for Si as well as multiphoton excitations [19]. Similarly, Gilbert et al. observed photoionization from fs laser pulses with low LPE in Si [20]. Photoionization is not expected to play a role for metallic materials since there is no bandgap and any ionization should be re-neutralized by one of the abundant and highly mobile free electrons [19]. Tamura et al. theoretically investigated the electric potentials during field evaporation of insulators with photoinduced holes [21]. In contrast to metals, the field is only partly screened at the surface and a gradient of the electric potential is present inside the material. This leads to accumulation of holes at the surface and in turn to a reduction of the energy barrier for desorption [21,22]. Chiaramonti et al. developed a setup for APT with extreme UV radiation via a high harmonic generation system based on noble-gas-filled capillaries [23]. Depending on the identity of the noble gas, wavelengths between 28 nm and 50 nm could be achieved [24]. These wavelengths correspond to photon energies of 25-45 eV. The introduced energy of a single pulse is too low to heat the apex of the tip significantly. The authors propose that a direct athermal photoionization with a thermal desorption might take place [23].

Equation 1 shows the evaporation rate $\Phi_{evap}(t)$ during thermally assisted field evaporation. It is dependent on the height of the energy barrier, which can be simplified written as a function of the electrostatic field $F(t)$ as $Q = Q_0 - \sqrt{\frac{F(t)}{F_0}}$. $T(t)$ is the absolute temperature, $v_0$ the vibration frequency, $k_B$ the Boltzmann constant and $F_0$ the evaporation field of the sample [4,25,26]. The terms are shown as being time-dependent since time-of-flight mass spectroscopy is used to identify the ion species. While voltage pulses reduce the energy barrier for evaporation due to the increase in electrostatic field, laser-pulses usually are understood to increase the temperature of the specimen to initiate evaporation. The energy barrier $Q_0$ for an ion to evaporate can be described by the energy difference between the adatom and the ion in vacuum, as shown by Müller [27].

$$\Phi_{evap}(t) = v_0 * exp\left\{-\frac{Q_0 - \sqrt{\frac{F(t)}{F_0}}}{k_B T(t)}\right\} \qquad (1)\ [4]$$

Equation 2 depicts the Fowler-Nordheim type equation, which describes the current of electrons tunneling through an energy barrier $Q_{FN}$ in an electrostatic field [28,29]. $A$ and $B$ are constants that include among other things the mass and charge of the electron. Historically, the equation was found for electron emission from metals, with the sample at negative potential. It was shown that the field is interchangeable with temperature as depicted in equation 3. The emission of thermionic electrons [30] at high temperatures is based on the same processes as emission of electron in a high electrostatic field [28,31]. The equation is, however, not limited to electron emission of metals, but describes in general the tunneling through a triangular energy barrier in the presence of an electrostatic field, which is widely applicable for instance in semiconductor devices [29,32].

$$j = A * \frac{F^2}{Q_{FN}} * exp\left\{-\frac{B * Q_{FN}^{3/2}}{F}\right\} \qquad (2)\ [28]$$

$$j_T \sim T^2 * exp\left\{-\frac{const.}{T}\right\} \qquad (3)\ [28]$$

APT specimens were investigated from a metallic alloy containing about 80 wt.% Ni and 20 wt.% Cr (Ni-20Cr) and thermally grown chromia ($Cr_2O_3$). A two-step electropolishing process with perchloric acid solutions was performed on Ni-20Cr wire to achieve sharp specimens [33]. The chromia was grown on a Ni-base alloy and contains about 3 at.% other elements, mostly Ti. Lift-out and sample preparation were performed from a cross-section of the oxide with the focused ion beam (FIB) of a dual-beam FEI Versa 3D workstation (Hillsboro, Or, USA) implementing well-known procedures [34,35]. It is the same material as in [36] and more details can be found there as well as in [37]. APT experiments were performed on a CAMCEA LEAP 6000 XR (Madison, Wi, USA) at 40 K test temperature, 200 kHz frequency and different LPEs and DRs. The tool has 52% detector efficiency and is equipped with a deep-UV laser system that provides laser pulses with 257.5 nm wavelength corresponding to 4.8 eV photon energy. The deformed top layer from FIB processing was removed in the APT in case of the chromia sample before the actual experiment. Experiments were performed at LPEs between 1 and 70 pJ. Details on the measurement procedure are described in the supplementary material. Additionally, the DR of chromia as a function of electrostatic field was investigated at different voltages between 4702 to 7160 V in steps of 200 V for experiments with 5 and 50 pJ, as well as at 40 and 80 K test temperatures, respectively. Measurement data was exported from the commercial software AP Suite 6.3 and processed and visualized in OriginPro 2023.

Figure 1 shows the DR as a function of LPE while the voltage was held intermittently constant for the metallic alloy Ni-20Cr. The tip shape and its blunting led to an increase in standing voltage from about 3450 to 3545 V, or 2.8 %, during the collection of little more than 1 million ions. The DR decreased slightly at high LPEs during the holding segments at constant voltage due to tip blunting. Therefore, only the initial 100 000 ions were selected for averaging in the case of 35 and 40 pJ. The DR, sample windows, LPE and voltage as a function of ion sequence number during the experiment can be found in the supplementary material in Fig. S1. The peak temperature during the pulse has been shown to be proportional to LPE for metals [13]. Thus, T can be substituted with LPE in equation 1. The Arrhenius fit, corresponding to the expected thermal evaporation, has an $R^2$ value of 0.9999 and the fit parameters can be found in Table S1 in the supplementary material.

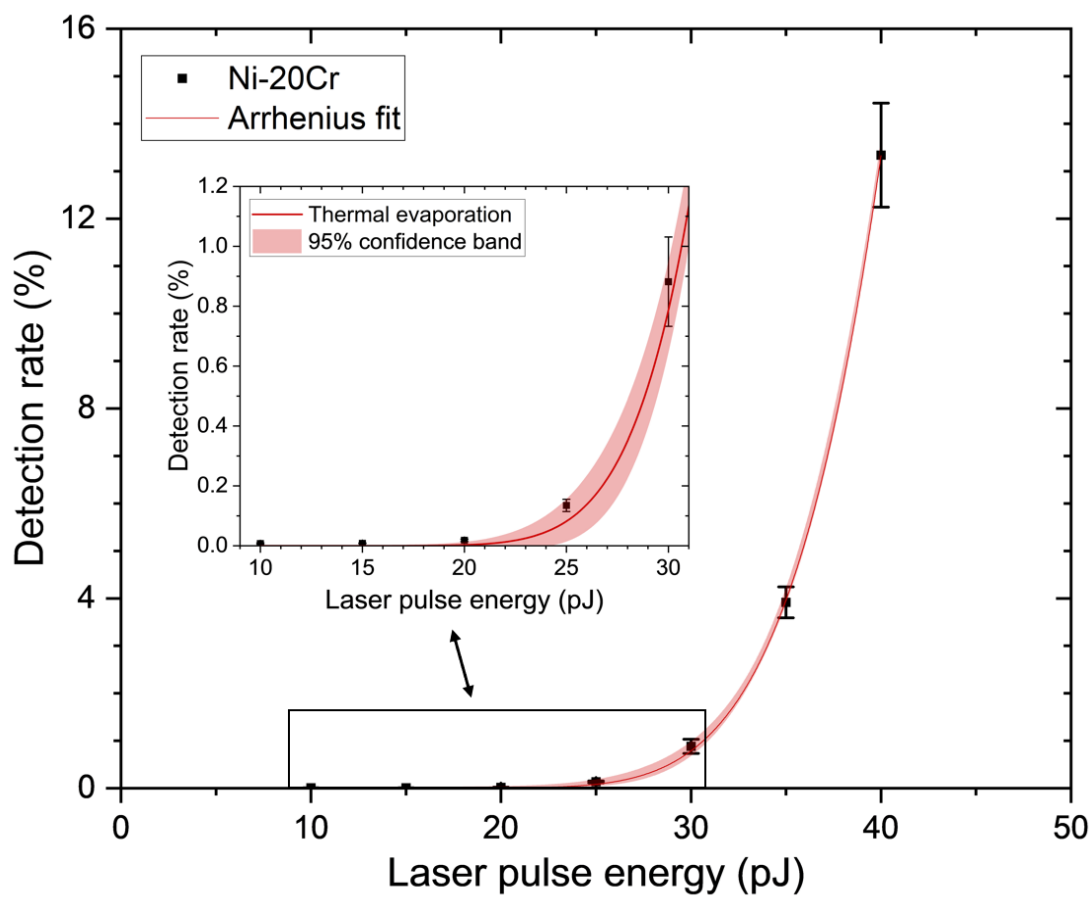

Figure 1: DR as a function of LPE during constant voltage experiments for the metallic alloy Ni 20Cr. The inset shows a detailed section of low LPEs. An Arrhenius fit is depicted as well.

Figure 2 shows the DR as a function of LPE at constant voltage for the thermally grown chromia specimen. The chromia specimen exhibits a small shank angle and hence shows minimal change in voltage of about 15 V during the entire experiment (see S2 in the supplementary as well as the discussion in [36]). At low LPEs the DR depicts a linear relationship with LPE. At higher LPEs, the data points significantly deviate from a linear fit. The evaporation does not simply match with an additional thermal contribution. However, the whole range of data points can be fitted to $R^2$=0.9999 with equation 4, where $x_1 = LPE$ and $x_2 = (T_0 + k*LPE)$. A, B, C and k are constants, and the values are shown in Table S2 in the supplementary material. The fit is depicted with the 95% confidence range in Fig. 2. An explanation for this observation is presented in the discussion section below.

$$A*x_1 + B*x_2^2 * exp\left\{-\frac{C}{x_2}\right\} \qquad (4)$$

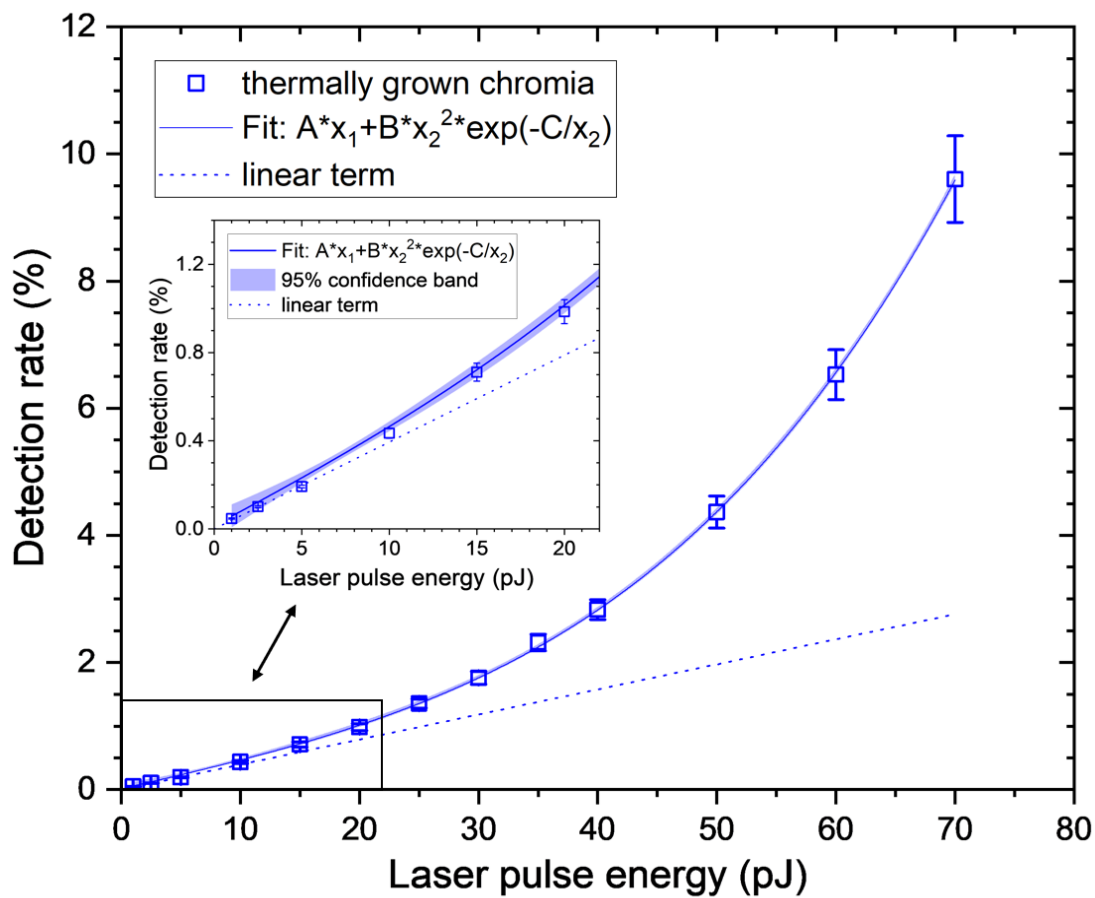

Figure 2: DR as a function of LPE for thermally grown chromia during constant voltage experiments. The solid line shows the apparent fit function with 95% confidence range and the dashed line shows the linear term of the fit. The inset provides a detailed depiction of low LPEs.

Figure 3 depicts the DR as a function of electrostatic field estimate for thermally grown chromia. Relative field changes are displayed since Kingham showed field calculations only for metallic samples [38]. Laser-assisted evaporation of $Fe_3O_4$ showed that the charge state ratio (CSR) of metallic Fe ions can depend on the LPE [39]. The CSR of Cr++/Cr+ ions during evaporation of chromia show a consistent trend as a function of LPE (shown in Fig. S3 in the supplementary) and hence can be used to compare the field strength between experiments. Data points from an initial 0.5 % target DR were used to put the experiments into relation to each other and all further datapoints are derived from the voltage steps. This means that the relative field estimate is accurate. See the supplementary material for the depiction of field estimate as a function of CSR. Fig. 3 shows four curves corresponding to four experiments with 5 and 50 pJ LPE as well as at 40 and 80 K test temperature. The parameters were chosen to show, on one hand, experiments where no significant thermal pulse is expected (5 pJ LPE) and, on the other hand, a high LPE, where one would expect a thermal contribution due to the laser pulse (50 pJ). All curves follow the Fowler-Nordheim type equation shown in equation 2 with $R^2>0.999$ and the fit parameters can be found in Table S3 in the supplementary material. Fig. 3 shows that for the low LPE the initial test temperature has a significant influence. The curves with 5 pJ are shifted to each other and at the same field estimate the experiment at 80 K shows more evaporation. The data for 50 pJ are shifted towards lower fields as expected. Furthermore, they exhibit a greatly reduced role of initial temperature compared to the curves with 5 pJ LPE. This fits with the assumption of a temperature increase during the laser pulse with high LPEs [14].

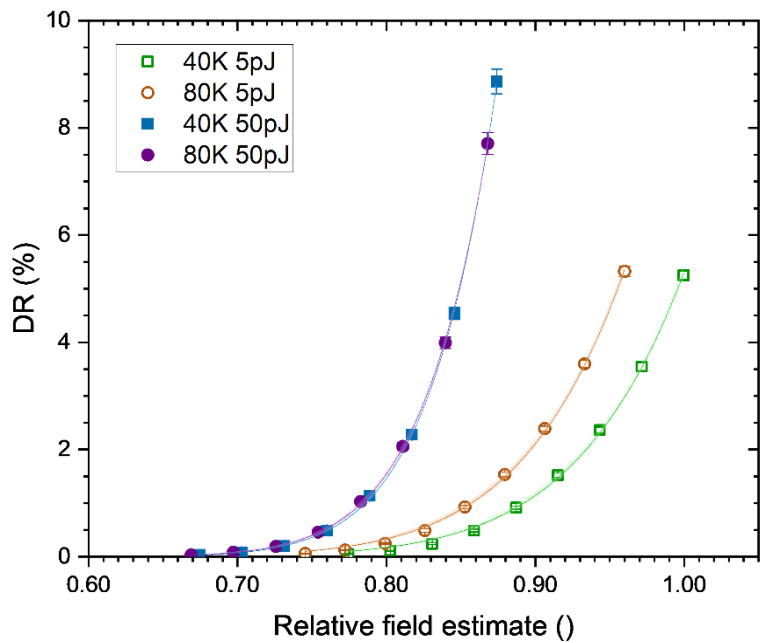

Figure 3: DR as a function of field estimate for thermally grown chromia during stepwise voltage experiments. The solid lines show the Fowler-Nordheim type fit functions with 95% confidence range. For further details see the text.

The experiment on the metallic alloy Ni-20Cr shows a strong correlation with the expected Arrhenius fit, i.e. thermal evaporation of ions. In contrast to that, the evaporation of thermally grown chromia by deep-UV laser pulses shows a linear correlation at low LPEs. This linear regime suggests that the evaporation mechanism involves direct photoionization since the DR is proportional to the LPE, i.e. the number of photons illuminating the specimen. The deviation from the linear regime is discussed below.

The presence of a fit to the Fowler-Nordheim type equation, as seen in Fig. 3, points to electron tunneling as the rate-limiting step in ion formation. Our observations suggest that the photon illumination in combination with the electrostatic field triggers electron tunneling and thereby creating holes. In semiconductors, the field inside the specimen is only partially screened and hence a negative electric potential gradient is present inside the material as well as down the shank of the needle [15,21,40]. Hence, the electrons can tunnel in the direction of the negative potential gradient leaving holes behind. In general, the absorption of laser light with higher energy than the bandgap of a semiconductor leads to the formation of electron-hole pairs. The high electrostatic field separates the electrons from the holes. These free charge carriers, i.e. holes, accumulate at the surface of the tip. So far, thermal evaporation was assumed during evaporation of semiconductors with lasers that have photon energies in the low eV range, where the holes absorb the laser energy and cause localized heating of APT specimens. With increasing photon energy, a higher surplus energy is being transferred to the lattice. In the case here, excess photons that are not directly involved in the ion generation still convert their energy to the heating of the APT specimen, especially at higher LPEs, and provide a higher probability for electron tunneling. The fit shown in equation 4 and Fig. 2 assumes that the temperature rise is proportional to the LPE. Silaeva et al. showed that the temperature increase during laser pulsing of semiconductors is dependent on the charge carrier density and the electrostatic field, and have in a first approximation two linear regimes [14]. The fit parameters for thermally grown chromia suggest a temperature rise of 46 K during a pulse with 70 pJ LPE. This moderate temperature increase should lie within the initial slope. Hence, the assumption of proportional temperature rise with LPE is validated. Combining the evaporation contributions, we can see that a superposition of photoionization with photofield tunneling of electrons fits with the observed dependence of DR with LPE for the measurement of thermally grown chromia with deep-UV laser light. At low LPE, photoionization, corresponding to pathway i) described by Tsong [19], dominates. With increasing contribution of heating the sample during the pulse, more electrons tunnel due to the photon illumination and the electrostatic field leaving holes behind. The desorption of an ion is apparently not rate-limiting and may happen by thermal vibrations as the energy barrier is decreased in the presence of holes [21,22].

The photon energy of the experiments presented here (257.5 nm wavelength, 4.8 eV photon energy) is higher than the bandgap of 2.9 to 3.55 eV of chromia [41,42]. Thus, photoionization should in principle be possible. Chromia was previously evaporated with UV laser pulses (355 nm wavelength, 3.49 eV photon energy) with LPEs between 15 and 115 pJ [43]. The O concentration was measured to be close to the expected stoichiometry of 60 at.%. The presence of other elements in thermally grown oxides should provide further energy states and hence reduce the bandgap, similar to doping. Furthermore, the strong electric field at the apex of the APT tip might compress the bands even more [15]. Therefore, the UV laser pulses with 3.49 eV photon energy should already be overcoming the bandgap of the material. However, the rather constant O concentration as a function of LPE could be explained by a high electrostatic field in combination with thermal pulses from the laser, as discussed in [36]. In contrast to that, deep-UV laser pulses (4.8 eV photon energy) enabled evaporation over more than three orders of magnitude between 0.03 pJ and 90 pJ. The O content was strongly dependent on the electrostatic field, presumably due to post-ionization of otherwise neutral O from dissociations [36].

The work function describes the minimum energy necessary to remove an electron from a material. Chromia has a work function of about 4.5 eV [44,45]. Since the photon energy of the deep-UV laser light used here (4.8 eV photon energy) is higher than the work function of chromia, a photon of the laser pulse has high enough energy to cause the removal of an electron from its atom due to the photoelectric effect. The value of the photon energy in comparison to the work function of the material is most likely a prerequisite for the observed photofield tunneling. Further investigations with different materials need to be performed to elucidate this.

A similar dependence of the O content on LPE for chromia shown in [36] has been observed by Devaraj et al. for UV (355 nm wavelength, 3.49 eV photon energy) laser assisted evaporation of magnesia (MgO) [46]. A similar wide range of possible LPEs was observed. The authors furthermore showed that the electric field, indicated by the CSR $Mg^{++}/Mg^{+}$, is correlated with LPE and O content in a similar manner. Magnesia has a higher bandgap (7.7 eV) than the photon energy (3.49 eV) of the used laser light. However, sub-bandgap states are expected at corner sites as well as in the high electric field [47]. The experimentally observed ion flux as a function of LPE and applied voltage in [22] fits well with the Fowler-Nordheim type equation, suggesting that the same mechanism is in play for magnesia as for chromia. Similarly, APT investigations of magnetite ($Fe_3O_4$) with UV laser light showed a wide range of possible LPEs [39]. In this study as well as for chromia, the voltage and electrostatic field is correlated with the LPE as $F \sim 1 - ln(LPE)$. This supports the assumption that the same mechanisms were in effect. No inversion of CSRs and significant counts of $O_2^+$ are observed for chromia, as was the case for magnetite. Thus, further investigations regarding the developing charge states should be carried out in the future.

**Conclusion**

In this work, the evaporation mechanism of thermally grown chromia during APT experiments with deep-UV laser light is investigated. Holding the APT microtip at a constant voltage allows to elucidate the dependence of DR on LPE as well as the magnitude of the electrostatic field. The metallic alloy Ni-20Cr was investigated in comparison and shows the expected Arrhenius law, i.e. thermal evaporation. In contrast to that, the DR of thermally grown chromia as a function of electrostatic field follows the Fowler-Nordheim type equation, speaking for photofield tunneling as the rate-limiting step during evaporation. In semiconductors the electrostatic field is only partly screened at the surface and a negative electric potential gradient is present within the specimen as well as down the shank. Electrons under laser illumination can seemingly tunnel towards the negative potential gradient leaving holes behind. The DR as a function of LPE shows a linear regime, corresponding to photoionization, at low LPEs. At higher LPEs, the DR follows a relationship represented by an additional term for electron tunneling. The behavior can be explained by excess photons heating the specimen and allowing more tunneling. The desorption of ions is not rate-limiting and may be thermal after reduction of the energy barrier by the presence of holes in the vicinity. The influence of bandgap and work function are discussed. More investigations are needed to confirm the role of work function for this evaporation behavior as well as to understand the emerging charge states.

**Supplementary material section**

There is a connected supplementary material file with additional information regarding the measurement procedure, tip blunting as well as field estimates. Furthermore, Tables with fit parameters are provided.

**Acknowledgements**

The experiments were performed at Chalmers Materials Analysis Laboratory (CMAL). The authors gratefully acknowledge Dr. Anton Chyrkin, Department of Chemistry, Chalmers University of Technology, for providing the chromia sample material and Dr. Andrea Fazi, previously at the Department of Physics, Chalmers University of Technology, for FIB sample preparation.

**Conflict of interest**

The authors have no conflicts to disclose.

**Data availability statement**

The data that support the findings of this study are available from the corresponding author upon reasonable request.

# Supplementary material

for article

# Ion generation by photoionization and photofield tunneling of electrons during atom probe tomography of thermally grown chromia with deep UV laser light

Severin Jakob[1], David Mayweg[1], Mattias Thuvander[1]

[1] Department of Physics, Chalmers University of Technology, SE-412 96 Göteborg, Sweden

To avoid significant influence of tip blunting during the evaporation at different LPEs, the following procedure was implemented: The voltage was initially stabilized during 20 or 30 pJ LPE and a target DR of 1 %. Then, the software feedback loop between DR and specimen voltage was turned off, i.e. the voltage was held constant. The LPE was changed, and the resulting DR was recorded for two minutes or about 200 000 ions. The LPE was switched back to 20 or 30 pJ and the voltage was adjusted to again get a target DR of 1%, before starting this procedure over for the next LPE. Fig. S1 and S2 show voltage, DR and LPE as a function of ion sequence for the metallic alloy Ni-20Cr and thermally grown chromia.

Tip blunting is not an issue for chromia as can be seen in Fig. S2. Furthermore, tip blunting was thoroughly discussed in [1] and an SEM image of a microtip can be found in the connected supplementary there.

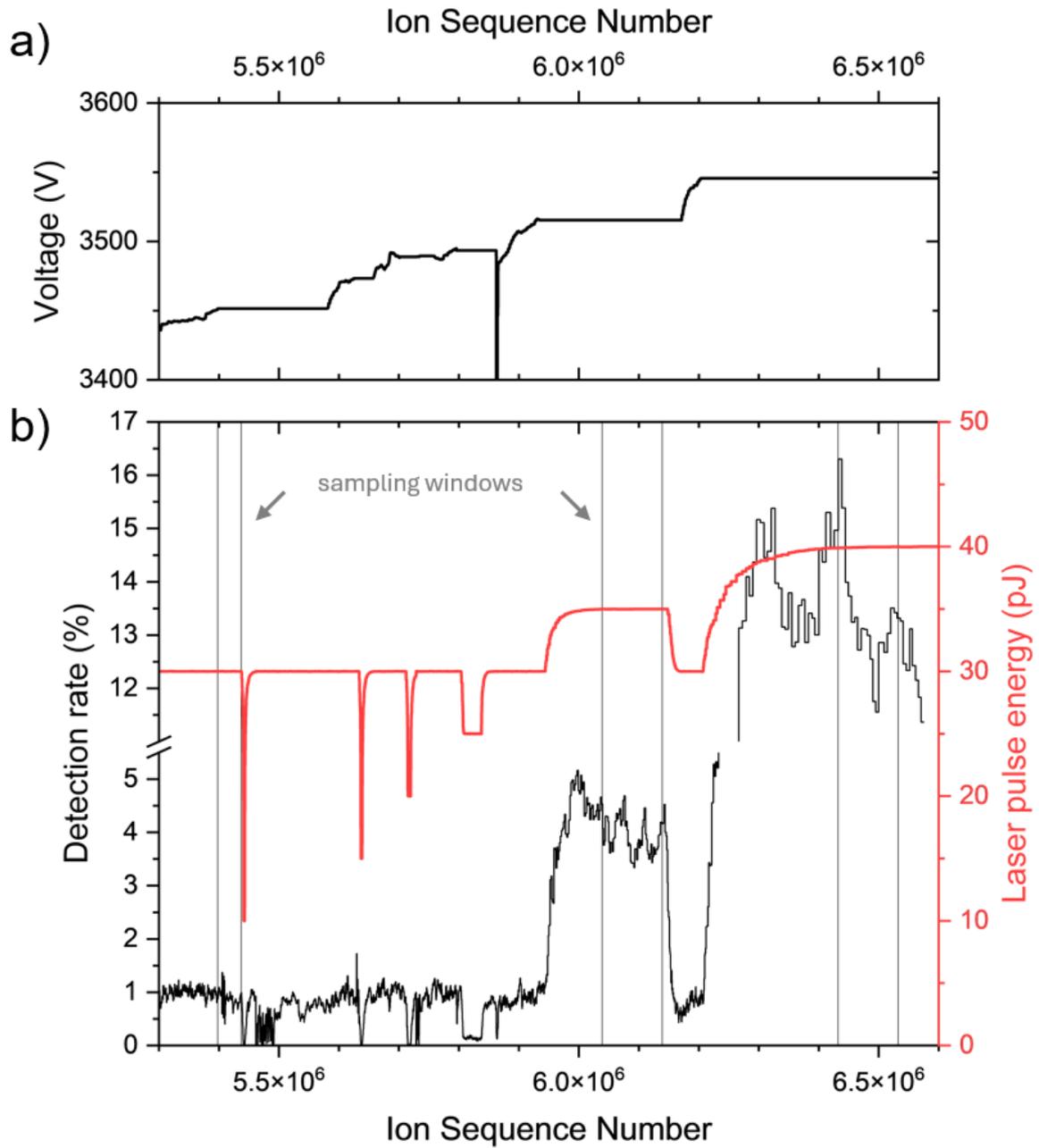

Figure S1: a) Voltage history as a function of ion sequence number during LPE variation with constant voltage segments; b) DR and LPE as a function of ion sequence number for the experiment on the metallic alloy Ni-20Cr. The sampling windows for 30 pJ as well as 35 pJ and 40 pJ are depicted in (b).

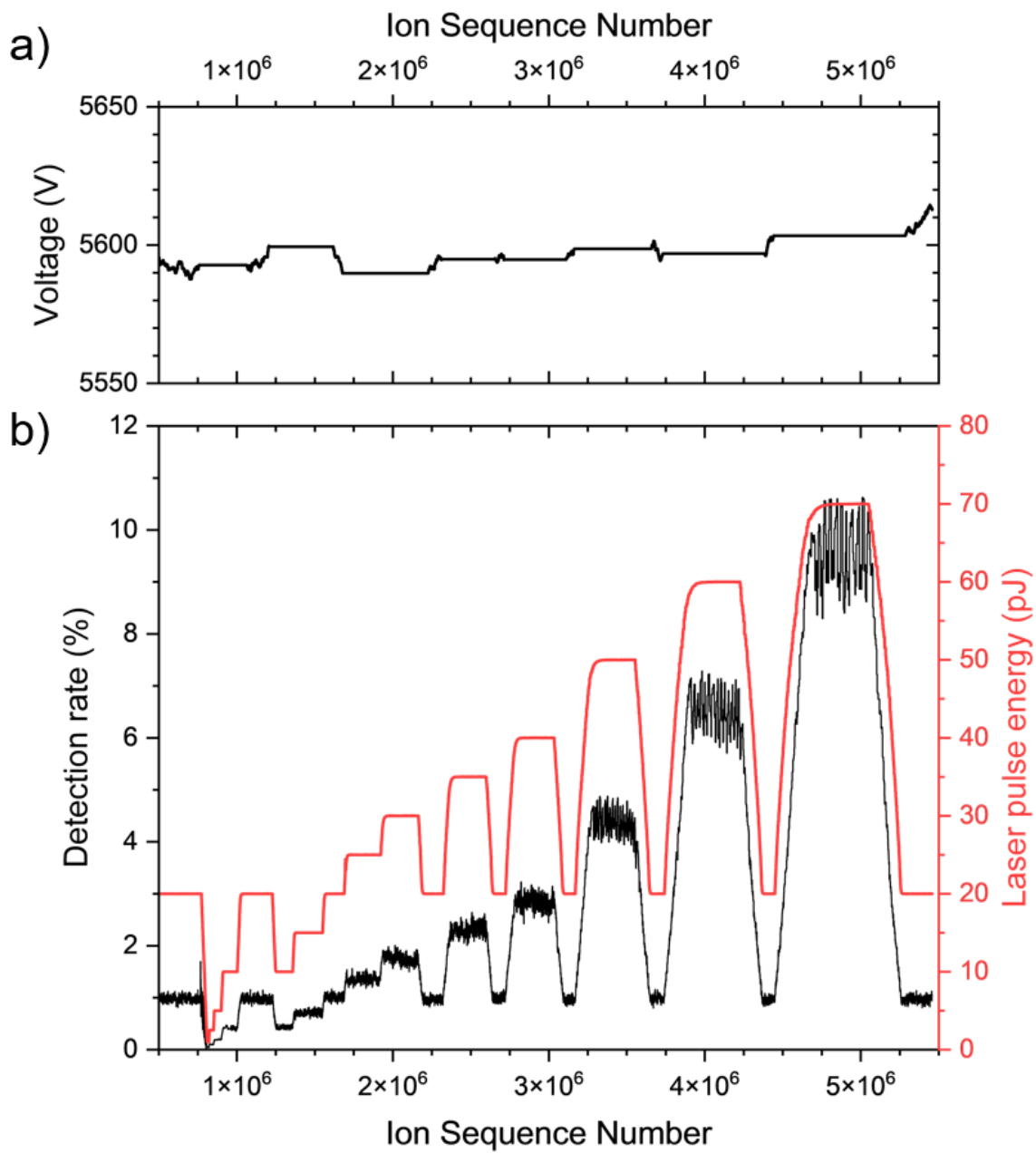

Figure S2: a) Voltage history as a function of ion sequence number during LPE variation with constant voltage segments; b) DR and LPE as a function of ion sequence number for the experiment on thermally grown chromia.

Experiments regarding the evaporation behavior as a function of electrostatic field were performed on the same microtip of thermally grown chromia at different voltages with 5 and 50 pJ LPE at 40 and 80 K test temperature, respectively. After establishing a target DR of 0.5 % the voltage was held constant and changed in steps of 200 V. The DR was recorded for 2 minutes before the voltage was changed to the next value.

Field estimates can be calculated by CSRs for metallic samples according to Kingham [2]. Laser-assisted evaporation of $Fe_3O_4$ showed peculiar dependence of CSR on the LPE [3]. In contrast to that, chromia shows a range of consistent decreasing CSR with increasing LPE as can be seen in Fig. S3(a). The lines indicate 5 and 50 pJ LPE, at which the stepwise voltage experiments were conducted. Fig. S3(b) shows the field estimate of data from the stepwise voltage experiments. The datapoints from the initial 0.5 % target DR are indicated with lines and were used to compare the experiments to each other. Field estimates were iterated to establish an overlapping range of CSR with the field as indicated by the linear trendline. At low LPE and high field, no significant Cr+ peak is visible and prevents a meaningful calculation of CSR.

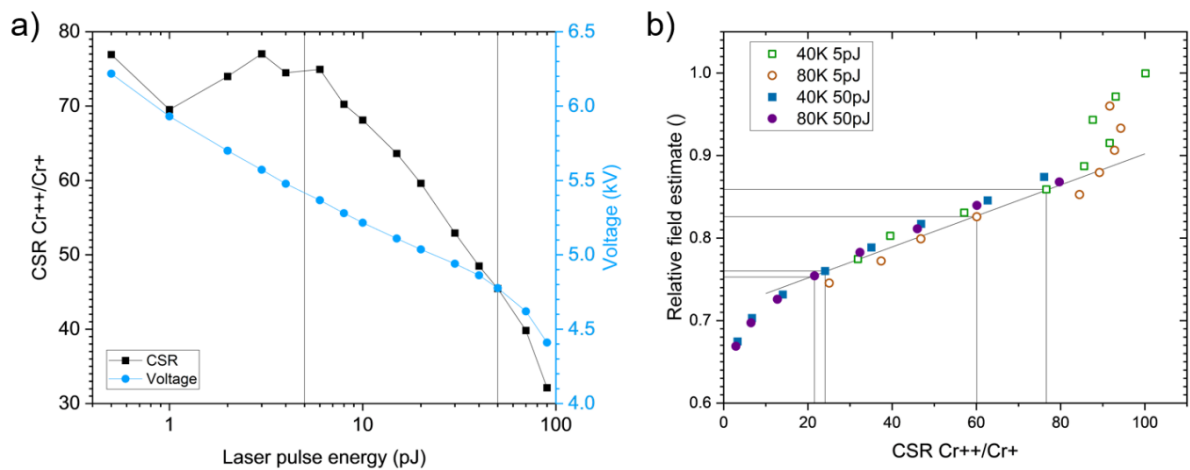

Figure S3: a) Voltage and Cr++/Cr+ CSR as a function of LPE in laser-assisted field evaporation of thermally grown chromia with a target DR of 0.5 %. These datapoints were measured during experiments in [1]. b) Relative field estimate for the experiments with stepwise voltage increments.

Table S1: Fit parameters for the relationship of DR as a function of LPE for the experiment on Ni20Cr.

| $A * exp\left\{-\frac{B}{LPE}\right\}$ | A | error | B | error | $R^2$ |
|---|---|---|---|---|---|
| | 6.46e4 | 6.0e3 | 340 | 3.8 | 0.9999 |

Table S2: Fit parameters for the relationship of DR as a function of LPE for the experiment on thermally grown chromia: $A * x_1 + B * x_2^2 * exp\left\{-\frac{C}{x_2}\right\}$, $x_1 = LPE$ and $x_2 = (T_0 + k * LPE)$.

| $R^2=0.9999$ | value | error |
|---|---|---|
| A | 0.039 | 0.008 |
| B | 0.049 | 0.098 |
| C | 341 | 18.2 |
| k | 0.656 | 0.368 |

Table S3: Fit parameters for the relationship of DR as a function of electrostatic field for experiments on thermally grown chromia.

| $A * F^2 * exp\left\{-\frac{B}{F}\right\}$ | A | error | B | error | $R^2$ |
|---|---|---|---|---|---|
| 40 K, 5 pJ | 1.21e3 | 3.3e2 | 308 | 7.0 | 0.9990 |
| 80 K, 5 pJ | 1.52e3 | 3.9e2 | 299 | 6.4 | 0.9991 |
| 40 K, 50 pJ | 6.98e5 | 1.0e5 | 395 | 3.4 | 0.9999 |
| 80 K, 50 pJ | 3.53e5 | 5.6e4 | 380 | 3.7 | 0.9999 |